\documentclass[aps,amssymb,amsmath,twocolumn,secnumarabic]{revtex4}
\usepackage[dvips]{graphicx}
\usepackage{amsfonts}
\usepackage{amsmath}
\usepackage{mathrsfs}

\usepackage[arrow, matrix, curve]{xy}
\xymatrixcolsep{0.5cm} \xymatrixrowsep{0.5cm}

\def\L2{{\it Fun\/}\bigl(\text{\GL}\bigr)}

\def\la{\lambda}

\newcommand{\rd}{\, \mathrm{d}}
\def\a{\alpha}
\def\da{{\dot{\alpha}}}
\def\la{\lambda}
\def\tla{{{\tilde{\phantom{a}}\hspace*{-2mm}\lambda}}}
\def\tmu{{\tilde{\mu}}}
\def\tchi{{\tilde\chi}}
\def\trho{{\tilde\rho}}

\def\bfp{\bar f^+}
\def\bfm{\bar f^-}
\def\fp{f^+}
\def\fm{f^-}

\begin{document}

\title{Towards a Worldsheet Description of ${\mathcal N}=8$ Supergravity}

\author{Arthur Lipstein$^1$ and Volker Schomerus$^2$}

\affiliation{$^1$ II. Inst. f\"ur Theoretische Physik, University of Hamburg,
Luruper Chaussee 149, D-22761 Hamburg,\\
$^2$ DESY Theory Group, DESY Hamburg, Notkestrasse 85, D-22607 Hamburg,
Germany}

\date{July 2015}

\begin{abstract}
In this note we address the worldsheet description of 4-dimensional ${\mathcal N}=8$ supergravity
using ambitwistors. After gauging an appropriate current algebra, we argue that the only physical vertex operators correspond to the ${\mathcal N}=8$ supermultiplet. It has previously been shown that worldsheet correlators give rise to supergravity tree level scattering amplitudes. We extend this work by proposing a definition for genus-one amplitudes that passes several consistency checks such as exhibiting modular invariance and reproducing the expected infrared behavior of 1-loop supergravity amplitudes. \\
\vskip 0.1cm \hskip -0.3cm
\end{abstract}

\maketitle

\vspace*{-8cm} \noindent
{\tt \phantom{DESY 15-nnn}}\\[5.2cm]

\section{Introduction}

Worldsheet descriptions of quantum field theories can uncover deep and
unexpected insights into conventional models of particle physics. While
the AdS/CFT correspondence provides a systematic framework to find a string
theoretic formulation, the precise relation with the particle theory is
somewhat obscured by strong/weak coupling duality.
This motivated a number of attempts to construct weak-weak coupling
dualities between field and string theory, primarily in the context of
twistor string models \cite{Witten:2003nn,Berkovits:2004hg}. Even though
most of these worldsheet models are
not equivalent to the field theory they were designed to describe, they
still became a valuable source for new ideas and results. For
10d supergravity theories, remarkable recent progress has
allowed for the calculation of tree and loop level scattering
amplitudes from a worldsheet description known as ambitwistor string theory \cite{Cachazo:2013hca,Mason:2013sva,Adamo:2013tsa,Geyer:2015bja}.

One of the most interesting gravitational theories, however, is
${\mathcal N}=8$ supergravity in $d=4$ space-time dimensions. It
is an important open question whether this theory is UV finite or
not \cite{Bern:2006kd}. If it were possible to recast $4d$ ${\mathcal N}=8$ supergravity
in the form of a string theory, finiteness might become manifest.
This is the context our work draws its motivation from. One might
hope that the recent developments for 10d supergravity could
provide a promising starting point. On the other hand, Green at al.
have shown that one can not decouple ${\mathcal N}=8$ supergravity
from the compactification of 10d superstring theory \cite{Green:2007zzb}.

We therefore pursue a different route here by using an ambitwistorial version of a model developed by David Skinner \cite{Skinner:2013xp} that was shown to reproduce the correct tree-level amplitudes of $\mathcal{N}=8$ supergravity in \cite{Geyer:2014fka}. Our main goal is to investigate loop amplitudes. Even though the ambitwistor model for 4d supergravity does not naively appear to be a
stringy theory because it contains an ungauged Virasoro symmetry, the
natural prescription we propose for the 1-loop amplitudes possesses
string-like features while, at the same time, reproducing central
properties of supergravity.

The plan of this note is as follows. In the next section we
will set up the general framework. Part of this simply reviews
constructions from \cite{Skinner:2013xp,Geyer:2014fka}. In addition we shall
define the notion of a physical vertex operator and argue that,
even after the inclusion of a Ramond sector, the only physical
vertex operators correspond to positive and negative helicity states
of the ${\mathcal N}=8$  supermultiplet. The 1-loop amplitude is then
defined and computed in section 3, and in section 4 we show that it is
modular invariant, satisfies spacetime momentum conservation
and reproduces the expected IR limits. We conclude
with a discussion of results and open problems.

\section{The Ambitwistor Theory}

\subsection{Ambitwistor fields}

The ambitwistor model consists of two sectors that we shall refer
to as the matter and ghost sector, respectively. The action of the matter
sector takes the form
$$ S_m \sim \int_\Sigma d^2z \left( W \cdot\bar \partial Z +
\trho \cdot \bar \partial \rho  \right)$$
where $Z$ and $W$ denote multiplets $Z=(\la_\a,\mu^{\da};\chi^a)$ and $W=(\tmu^\a,\tla_{\da};\tchi_a)$ with $\a,\da =1,2$ and $a=1,\dots,8$.
The first four components are bosonic while the remaining eight are
fermionic. They form four $\beta\gamma$ and eight $bc$ systems of
conformal weight $h_Z = 1/2 = h_W$, respectively. The second pair of
multiplets $\rho=(\rho_\a,\rho^\da;\omega^a)$ and $\trho = (\trho^\a,\trho_\da;\tilde{\omega}_a)$ in the matter sector is
similar to the $WZ$-system except that all gradings are reversed, i.e.\
the first four components are fermionic while the remaining eight are
bosonic. Let us note that the total central charge of these two sets
of multiplets is $c=0$.

In order to spell out the vertex operators of the ambitwistor model, we
must bosonize some of the $\beta\gamma$ systems in the matter sector.
To be more precise, let us introduce the free bosonic fields $\phi_\la$
and $\phi_\mu$ such that $\partial \phi_\la = \tmu^\a \la_\a$ and
$\partial \phi_\mu = \mu^\da \tla_\da$. Following the standard rules,
see e.g. \cite{Polchinski:1998rr}, their exponentials $\exp(s_\la
\phi_\la + s_\mu \phi_\mu)$ possess conformal weight $h_{\vec s} =
- s_\la^2 -s_\mu^2$. We shall consider operators of the form
$$\Phi = \varphi(W,Z,\rho,\trho) e^{s_\la \phi_\la + s_\mu \phi_\mu},
$$
where $\varphi$ is any expression composed from components of the
arguments and derivatives thereof, and refer to these as {\em operators in
the $(s_\la,s_\mu)$ sector} of the theory. One family that will play an
important role below is given by
\begin{equation}\label{mattervertex}
\tilde{{\mathcal H}}(\la,\tla;\eta):=\!\int\frac{\rd t }{t^3} e^{i t
\left(\tmu^\a(z)\la_\a + \tchi_a(z)\eta^a\right)}\delta^2(\tla - t \tla(z))
\end{equation}
which are parametrized by the bosonic variables $\la,\tla$ along with the
fermionic variables $\eta$. In order to clearly distinguish the parameters
from the fields, we have displayed the dependence on the worldsheet
coordinate $z$ in this formula. Using the standard identification
$\delta^2(\tla) \sim \exp(-\phi_\mu)$ we conclude that the fields $\tilde
{\mathcal H}$ belong to the $(s_\la,s_\mu)=(0,-1)$ sector of the theory.
It is not difficult to check that the integral over $t$ projects onto the
component of conformal weight is $h_{\tilde{\mathcal H}}= 1$.

Using the state-field correspondence, the ground state $|0,-1\rangle
\sim \exp (-\phi_\mu(0))|\text{vac}\rangle$ of the $(0,-1)$ sector satisfies the
following modified vacuum conditions:
\begin{equation} \label{vac}
\mu^\da_{n+1} |0,-1\rangle = 0 \quad  \quad \tla_{\da,n-1} |0,-1\rangle = 0
\end{equation}
for  $0< n \in \mathbb{Z}+\frac12$, and is annihilated by $K_n, 0<n \in
\mathbb{Z}$ for all other matter fields. This means that the modes $\mu^\da
_{1/2}$ are creation operators even though they lower the conformal weight,
while $\tla_{\da,-1/2}$ are annihilation operators in the $(0,-1)$ sector.
Such a partial swap of creation and annihilation operators is a crucial
ingredient in the oscillator construction of particle multiplets, see  e.g.
\cite{Gunaydin:1984vz} for the case of ${\mathcal N}=8$ supergravity. It is the key
mechanism by which the 4d ambitwistor model accommodates the ${\mathcal N}=8$
supergravity multiplet.

\subsection{The BRST operator}

The 4d model proposed by David Skinner in \cite{Skinner:2013xp} gauges a current algebra consisting of four
bosonic currents
\begin{eqnarray}
g & = & :Z \cdot W: \quad , \quad h = :\rho\cdot\trho: \\[2mm]
e^+ & = &  \langle \rho,\rho\rangle \quad \quad ,\quad
e^- = [\trho,\trho]
\end{eqnarray}
and four fermionic currents
\begin{eqnarray}
\fp & = & \langle Z,\rho\rangle \quad, \quad
\fm = Z \cdot \trho \\[2mm]
\bfp & = & \rho \cdot W
 \quad , \quad
\bfm = [W, \trho] \ \ .
\end{eqnarray}
The brackets $\langle , \rangle$ and $[ , ]$ instruct us to project
to the components carrying an undotted index $\a$ and a dotted index
$\da$, respectively, and to contract the indices with an $\epsilon$
tensor. It is easy to check that these currents form a $GL(1|1) \ltimes
\mathbb{R}^{2|2}$ current algebra at level $k=2$. The latter may be
obtained as a contraction of an $SL(1|2)$ current algebra.

Now we turn to the ghost sector, which contains two multiplets $C =
(C_A)=(c_g,c_h,c_+,c_-;\gamma_+,\gamma_-,\bar \gamma_+,\bar \gamma_-)$ and
$B = (B^A)$ accordingly. Note that the index $A$ runs over our basis
of currents. All eight components of the multiplet $C$ have conformal
weight $h_C=0$ while those in $B$ possess $h_B=1$. Following the
usual bosonization of ghost systems, we shall introduce two bosonic
fields $\phi$ and $\overline{\phi}$ such that $\partial \phi =
\beta^+\gamma_+ + \beta^-\gamma_-$ and similarly for $\partial
\overline{\phi}$. An operator is said to be in  the $(p,\bar p)$ picture
if it contains the factor
$$ e^{p\phi + \bar p \overline{\phi}} \quad \mbox{with} \quad
h_{\vec{p}} =  -p(p+1) - \bar p(\bar p+1)\ .  $$
From the fields in the ghost sector we can build a $GL(1|1) \ltimes
\mathbb{R}^{2|2}$ current algebra at level $k=-2$ using the standard
prescription and ultimately the usual BRST current $Q$. For the theory
to be anomaly free and nilpotency of the zero mode $Q_0$, it is crucial
to have $\mathcal{N}=8$ supersymmetry.

There has been some discussion whether the $c=0$ stress tensor of the
ambitwistor theory is gauged or not. We have verified that the combined
stress tensor of the ghost and matter sectors is non-trivial in the
cohomology of the current algebra BRST operator $Q_0$. One way out could
be to add further ghosts for the Virasoro algebra of the model. But this
would require to add additional matter fields with $c=26$, which appear
rather artificial. We believe that the Virasoro symmetry of the ambitwistor
model should be considered accidental, just as the transverse Virasoro
algebra that appears for flat space string theory in light-cone
gauge. Consequently, we suggest not to gauge the stress tensor of the
ambitwistor model.

\subsection{Physical vertex operators}

We can now specify what we mean by physical
vertex operators of the ambitwistor model. By definition, these
are scale-invariant integrated vertex operators in the cohomology
of this BRST operator $Q_0$. The integration is over the chiral
worldsheet coordinate $z$ so that the field in the integrand must
have conformal weight $h=1$.

It is not too difficult to check that the following vertex operators
in the $(-1,0)$ picture are physical:
\begin{equation} \label{VOmonepicture}
\tilde {\mathcal V}^{(-1,0)}(\la,\tla;\eta) = \int \rd z \delta(\gamma_+)
\delta(\gamma_-)\tilde {\mathcal H}(\la,\tla;\eta)\ .
\end{equation}
We shall refer to the integrand of this vertex operator as $\tilde {\mathcal V}^{(-1,0)}(\la,\tla;\eta)(z)$. In order to read off the picture, we recall that the ghost operator $\delta(\gamma_+)\delta(\gamma_-) \sim \exp(-\phi)$. The physical
vertex operators \eqref{VOmonepicture} turns out to describe negative
helicity states. Similarly, one can also construct physical vertex
operators ${\mathcal V}$ for positive helicity states. In the $(0,-1)$
picture they take a very similar form and are essentially obtained by complex conjugation. After performing complex conjugation, the vertex operators will depend on the fermionic variables $\tilde{\eta}$. For computational purposes however, it is convenient to Fourier transform $\tilde{\eta}$ into $\eta$ so that, see \cite{Geyer:2014fka} for more details,
\[
\mathcal{V}^{(0,-1)}(\la,\tla;\eta)=\int\rd z\delta(\bar{\gamma}_{+})\delta(\bar{\gamma}_{-})\mathcal{H}(\la,\tla;\eta),
\]
where
\[
\mathcal{H}(\la,\tla;\eta):=\!\int\frac{\rd t}{t^{3}}e^{it\mu^{\a}(z)\tla_{\a}}\delta^{2|8}\left(\lambda-t\lambda(z)|\eta-t\chi(z)\right).
\]
To obtain the physical vertex operators of positive and negative helicity
states in other pictures, we can apply the following picture-changing
operators
\begin{equation}
X = \delta(\beta^+)\delta(\beta^-) \fp \fm \, , \,
\overline{X} = \delta(\bar \beta^+) \delta(\bar \beta^-) \bfp\bfm\ .
\end{equation}
In particular, the integrand of a physical vertex operator for negative
helicity states in the $(0,0)$ picture is given by
\begin{eqnarray}\label{VOzeropicture}
& & \tilde {\mathcal V}^{(0,0)}(\la,\tla;\eta)(z)  =
\left(X(w) \tilde {\mathcal V}^{(-1,0)}
(\la,\tla;\eta)(z)\right)_{w\rightarrow z}  \nonumber \\
& & \quad = 
\left[\left\langle Z,\frac{\partial\tilde{\mathcal{H}}}{\partial W}\right\rangle +\left\langle \rho,\frac{\partial}{\partial W}\right\rangle \tilde{\rho}\cdot\frac{\partial\tilde{\mathcal{H}}}{\partial W}\right](\la,\tla;\eta)(z).  
\nonumber \end{eqnarray}
Similarly, application of the picture changing operator $\overline X$ to
${\mathcal V}^{(0,-1)}$ gives the vertex operator ${\mathcal V}^{(0,0)}$
for positive helicity states in the $(0,0)$ picture.

In order to argue that the ambitwistor model contains no further
physical vertex operators for particle multiplets, we can employ the
relation with oscillator representations (see comment above) to
show that the $\vec{s}=(0,-1)$ and $(1,0)$ sectors are the only
ones that can accommodate infinite multiplets including arbitrary
derivatives of the space time fields. In a second step one determines
the cohomology of the BRST operator $Q_0$ in the $L_0=1$ subspace of
these two sectors. Below we shall suggest that the ambitwistor theory
should also include a Ramond sector in which the $\trho \rho$ system
possesses anti-periodic boundary conditions. We claim that such a
Ramond sector does not give rise to additional physical vertex operators
either. In this sense, the ambitwistor theory contains no more than the
${\mathcal N}=8$ supergravity multiplet. A more detailed derivation of
these statements will be given in a forthcoming paper.

\section{Scattering Amplitudes}

Let us now turn to a discussion of scattering amplitudes. Since we have
not gauged the Virasoro symmetry of the model, we cannot employ the usual
formulae for amplitudes which involve Virasoro $b$-ghosts and Beltrami
differentials. The prescriptions we shall discuss here look very natural,
though they are justified only a posteriori when we establish the relation
with supergravity amplitudes.

In order to have non-vanishing amplitudes on a surface of genus $g$, the
total picture charges of all inserted vertex operators should add up to
$(p,\bar p) = (g-1,g-1)$. We have spelled out vertex operators in the pictures
$(0,-1), (-1,0)$ and $(0,0)$. On the sphere we can insert one operator in
the $(0,-1)$ and $(-1,0)$ picture each and then put all others in the $(0,0)$
picture. For the torus all vertex operators may be put in the $(0,0)$ picture.
In the following we will simply refer to such vertex operators as $\mathcal{V}$
and $\tilde{\mathcal{V}}$. Tree-level amplitudes were computed in \cite{Geyer:2014fka}
and they were shown to coincide with the $n$-point  N$^{k-2}$MHV amplitudes of
${\mathcal N}=8$ supergravity in four dimensions. Let us point out that, contrary
to the impression that is given in \cite{Geyer:2014fka}, the computations do not
involve a ghost sector for the Virasoro algebra of the ambitwistor theory.

Our main goal in this section is to propose an expression for the
1-loop amplitude of the ambitwistor theory. To this end we take the $(Z,W)$
to possess periodic boundary conditions, while $(\rho,\tilde{\rho})$ can be
either periodic or antiperiodic. The 1-loop amplitudes will then involve a
sum over spin structures as well as a GSO projection. For even spin structure,
the 1-loop $n$-point N$^{k-2}$MHV amplitude is given by
\[
\mathcal{A}_{even}^{(1)}=\int \rd \tau\Pi_{i=1}^{n} \rd z_{i}\left\langle \Pi_{l=1}^{k}\tilde{\mathcal{V}}_{l}\Pi_{r=k+1}^{n}\mathcal{V}_{r}\right\rangle ,
\]
where $\tau$ is the complex structure of the torus, and there is an implicit
integral over the zero modes of the $b,c$ ghost zero modes. Combining the
exponentials of the vertex operators with the action and integrating
out $(\mu,\tilde{\mu},\tchi)$ implies the equations of motion
\begin{eqnarray}
 \bar{\partial}\lambda &=&  \sum_{l=1}^{k}t_{l}\lambda_{l}\delta\left(z-z_{l}\right)     \nonumber \\
  \bar{\partial}\tla &=& \sum_{r=k+1}^{n}t_{r}\tla_{r}\delta\left(z-z_{r}\right) \nonumber \\
  \bar{\partial}\chi &=&   \sum_{l=1}^{k}t_{l}\eta_{l}\delta\left(z-z_{l}\right)\ .
\end{eqnarray}
For a genus-one worldsheet, the solutions are
\begin{eqnarray}
 \lambda(z) &=& \lambda_{0}+\sum_{l=1}^{k}t_{l}\lambda_{l}S_{1}\left(z-z_{l},\tau\right)  \label{lambda}    \\
\tla(z) &=&\tla_{0}+\sum_{r=k+1}^{n}t_{r}\tla_{r}S_{1}\left(z-z_{r},\tau\right) \label{tilde}  \\
  \chi(z) &=& \eta_{0}+\sum_{l=1}^{k}t_{l}\lambda_{l}S_{1}\left(z-z_{l},\tau\right)
\end{eqnarray}
where $S_{1}(z,\tau)$ is the Green's function on a torus with periodic
boundary conditions, and $(\lambda_{0},\tla_{0},\eta_{0})$
are zero modes. Note that there appear three other types of Green's functions,
which correspond to having antiperiodic boundary conditions along
at least one direction of the torus. We denote them by $S_{2,3,4}(z,\tau)$.
When these solutions are plugged into the delta functions of the vertex
operators, this gives rise to the 4d genus $g=1$ scattering
equations refined by helicity. Let us note that solutions to the
equations of motion can only exist if the sum of residues vanishes,
\[
R_{\lambda}=R_{\tla}=R_{\chi}=0\ ,
\]
where
\[
R_{\lambda}=\sum_{l=1}^{k}t_{l}\lambda_{l},\,\,\, R_{\tla}=\sum_{r=k+1}^{n}t_{r}\tla_{r},\,\,\, R_{\chi}=\sum_{l=1}^{k} t_{l}\eta_{l}.
\]
The remaining contributions to the correlation function arise from
integrating over fluctuations, which gives a 1-loop partition function,
and evaluating contractions of the $\rho,\tilde{\rho}$ fields, which
gives a determinant. The one-loop partition function can be composed
from the partition function of the components through standard
formulas. In the end, one obtains
\begin{multline}
\mathcal{A}_{even}^{(1)}=\int \rd \tau\frac{\rd^{2}\lambda_{0}\rd^{2}\tla_{0}
\rd^{8}\eta_{0}}{\mathrm{GL}(1)}\Pi_{i=1}^{n}\frac{\rd z_{i} \rd t_{i}}{t_{i}^{3}}
\\
\delta^{2}\left(R_{\lambda}\right)\delta^{2}\left(R_{\tla}\right)\delta^{8}
\left(R_{\chi}\right)\left(\Pi_{l=1}^{k}\delta_{l}\right)\left(\Pi_{r=k+1}^{n}
\delta_{r}\right)M\label{eq:1loopeven}
\end{multline}
where
\begin{eqnarray}
 \delta_{l} &=& \delta^{2}\left(\tla_{l}-t_{l}\tla\left(z_{l}\right)\right)
  \label{dl} \\
\delta_{r} &=&\delta^{2|8}\left(\lambda_{r}-t_{r}\lambda\left(z_{r}\right)|\eta_{r}-t_{r}
\chi\left(z_{r}\right)\right) \label{dr}  \\
  M &=&\sum_{\alpha=2,3,4}(-1)^{\alpha}\det H_{\alpha}\left(\theta_{\alpha}(0,\tau)/\eta(\tau)^{3}\right)^{-4} \label{M}\ .
\end{eqnarray}
Here $H_{\alpha}$ is an $n\times n$ block-diagonal matrix whose nonzero elements are
given by
\begin{eqnarray}
H_{\alpha}^{lm} &=& t_{l}t_{m}\left\langle \la_l,\la_m\right\rangle S_{\alpha}\left(z_{lm},\tau\right),\,\,\, l\neq m  \nonumber \\
H_{\alpha}^{ll} &=& t_{l}\left\langle \lambda\left(z_{l}\right),\la_l\right\rangle
\end{eqnarray}
for $ l,m\in\left\{ 1,...,k\right\}$, and
\begin{eqnarray}
H_{\alpha}^{rs} &=& t_{r}t_{s}\left[\tla_r,\tla_s\right]S_{\alpha}\left(z_{rs},\tau\right),\,\,\, r\neq s  \nonumber \\
H_{\alpha}^{rr} &=& t_{r}\left[\tla\left(z_{r}\right),\tla_{r}\right]
\end{eqnarray}
for $r,s\in\left\{ k+1,...,n\right\}$. The 1-loop amplitude for odd spin
structure can be computed in a similar manner. We will not need the form
of these contributions and therefore refrain from giving any details here.

\section{Consistency Checks}
In this section, we will describe various consistency checks for the amplitudes
computed in the previous section. For concreteness, we will focus on the formula
for 1-loop amplitudes with even spin structure in eq.\ \eqref{eq:1loopeven}. First
note that the amplitude is modular invariant, which is
suggestive of an interpretation as a string theory amplitude. This can be seen by noting
that if $\tau \rightarrow -1/\tau$, then $z\rightarrow z/\tau$, $t \rightarrow t/\sqrt{\tau}$,
$(Z,W) \rightarrow \sqrt{t} (Z,W)$, and $\theta_\alpha/\eta^3\rightarrow \tau^{-1}
\left(\theta_{\alpha}/\eta^{3}\right)$. Furthermore, all of these terms are invariant
under $\tau\rightarrow \tau+1$. As a result, we integrate $\tau$ over the fundamental
domain of the complex plane.

Next we verify that the amplitude encodes momentum conservation. To
see this, consider the sum of the momenta of the positive helicity
particles
\[
\sum_{r=k+1}^{n}\lambda_{r}\tla_{r}=
\sum_{r=k+1}^{n}t_{r}\tla_{r}\left(\lambda_{0}+
\sum_{l=1}^{k}t_{l}\lambda_{l}S_{1}\left(z_{rl},\tau\right)\right).
\]
To obtain the first equality, we used the delta functions in eq.\ \eqref{dl}
to replace $\tla_{r}$ with $t_{r}\tla\left(z_{r}\right)$
and then employed eq.\ \eqref{tilde}. Noting that
\[
\sum_{r=k+1}^{n}t_{r}\tla_{r}\sum_{l=1}^{k}t_{l}
\lambda_{l}S_{1}\left(z_{rl}\right)=
-\sum_{l=1}^{k}t_{l}\lambda_{l}\sum_{r=k+1}^{n}t_{r}
\tla_{r}\tilde{S}\left(z_{lr}\right)
\]
and inserting eqs.\ \eqref{lambda} and \eqref{dr}, one finds that
\[
\sum_{i=1}^{n}\lambda_{i}\tla_{i}=
\lambda_{0}\sum_{r=k+1}^{n}t_{r}\tla_{r}+
\tla{}_{0}\sum_{l=1}^{k}t_{l}\lambda_{l}.
\]
The right-hand-side vanishes on the support of the delta functions
$\delta\left(R_{\lambda}\right)$ and $\delta\left(R_{\tla}\right)$,
so momentum is conserved. Similar considerations apply for supermomentum
conservation.

In summary, the bosonic delta functions in eq.\ \eqref{eq:1loopeven} provide
$2n$ constraints on the integration variables, since four of the delta
functions ultimately encode momentum conservation. On the other hand,
there are $2n+4$ bosonic integrals, so in the end there are four
remaining integrals which can be interpreted as an integral over loop
momentum. In particular, after the delta functions localize the $2n$
coordinates of the vertex operators, one is left with an integral
over the zero modes $(\lambda_{0},\tla_{0})$,
and $\tau$. If we make the change of variables
$\tau=\log q$, the resulting integral is similar to one that would
arise from on-shell diagrams \cite{ArkaniHamed:2012nw}, where $q$
would be interpreted as a BCFW shift. In the present context, $q$
is related to the complex structure of the worldsheet, and the
integral over $q$ contains a UV cutoff due to modular invariance.

As a final consistency check, we verify that the amplitudes exhibit
the expected IR divergences, using techniques developed in
\cite{Adamo:2013tsa,Casali:2014hfa,Lipstein:2015rxa}. In the IR limit,
$\Im\tau\rightarrow\infty$ and the worldsheet degenerates
to a sphere with two additional punctures which are associated with
vertex operators with equal and opposite
momentum $k=\lambda_{0}\tla_{0}$. We will denote the additional vertex
operators as $\tilde{\mathcal{V}}_a,\mathcal{V}_b$. In this limit, the
contribution from odd spin structure vanishes,
and the sum of determinants in eq.\ \eqref{M} simplifies substantially.
Noting that $\eta(\tau)\rightarrow q^{1/24}$ and
\[
\theta_{\alpha}(0,\tau)\rightarrow\left\{ \begin{array}{c}
q^{1/8},\,\,\,\alpha=2\\
1,\,\,\,\alpha=3,4
\end{array}\right.
\]
where $q=e^{2i\pi\tau}$, one finds that
\[
M\rightarrow\det H_{2}+q\left(\det H_{4}-\det H_{3}\right).
\]
Furthermore, since
\[
S_{\alpha}(z,\tau)\rightarrow\left\{ \begin{array}{c}
1/z,\,\,\,\alpha=2\\
\mbox{constant},\,\,\,\alpha=3,4
\end{array}\right.
\]
we see that the contributions from $\alpha=3,4$ cancel out and $M\rightarrow\det H_{2}$,
which is the same determinant appearing in tree-level supergravity
amplitudes \cite{Geyer:2014fka}. Hence, in the IR limit a genus-one amplitude reduces
to an $(n+2)$-point tree level amplitude integrated over the on-shell
loop momentum $k$. Moreover, the IR divergent part of the amplitude
comes from the region of integration $k\rightarrow0$. In this limit, we can Taylor
expand $\tilde{\mathcal{V}}_a, \mathcal{V}_b$ in the soft momentum and keep the leading
order terms. The IR divergent part of the loop integrand can then be determined by using
Stokes theorem to express the soft vertex operators as contour integrals, integrating
them around each pair of hard vertex operators, and adding up the
residues. In the end, we obtain
\[
\mathcal{A}_{n}^{(1)}|_{div}=\int \rd^{4}k\delta\left(k^{2}\right)\sum_{i,j}\frac{\left(\epsilon\cdot k_{i}\right)^{2}}{k\cdot k_{i}}\frac{\left(\epsilon^{*}\cdot k_{j}\right)^{2}}{k\cdot k_{j}}\mathcal{A}_{n}^{(0)},
\]
where $\epsilon^{\alpha\dot{\beta}}=\xi^{\alpha}\tla_{0}^{\dot{\beta}}/\left\langle \xi\lambda_{0}\right\rangle$ and $\xi$ is a reference spinor. Evaluating the this integral using dimensional regularization then gives the standard 1-loop IR divergences of $\mathcal{N}=8$ supergravity \cite{Naculich:2011ry}.

\section{Conclusion and Outlook}
Above we argued that the spectrum of the critical, non-anomalous 4d ambitwistor model corresponds precisely to $\mathcal{N}=8$ supergravity, and define a prescription for computing loop amplitudes in this model. In doing so, it is not necessary to gauge the Virasoro symmetry as one does for the 10d ambitwistor string or the standard RNS string, since scale invariance and current
algebra symmetry of the world sheet theory appear to be powerful enough to remove all unphysical states. Hence, the Virasoro symmetry can be thought of as an accidental symmetry which does not impose further constraints on the spectrum. Recent studies of gravitational soft theorems using 4d ambitwistor model also suggest that Virasoro symmetry does not play an essential role as a symmetry of the gravitational S-matrix \cite{Lipstein:2015rxa}, so it would be interesting to make this connection more concrete.

We have also proposed a new formula describing 1-loop amplitudes of $\mathcal{N}=8$ supergravity which appears to have a very different structure than the 1-loop amplitude of 10d ambitwistor string theory.  For example, in the 10d formula, there is an additional delta function which enforces that $P_\mu P ^\mu =0$ vanish at some point on the worldsheet, where $P^\mu$ is the canoncial momentum of the worldsheet theory. This constraint ensures that only massless states can appear in factorization channels, but there is no such delta function in the 4d formula since this constraint is already built into the formalism. Furthermore, whereas in the 10d formula the delta functions determine the modular parameter $\tau$, in the 4d formula the delta functions leave $\tau$ unfixed giving rise to a 4d loop integral which is reminiscent of the formulae one obtains for 1-loop amplitudes of $\mathcal{N}=4$ super-Yang-Mills theory using on-shell diagrams. It is intriguing to see such structure emerging in the context of 4d gravitational amplitudes.

When computed on the torus, the loop amplitudes of ambitwistor models depend on elliptic functions and their equivalence to field theory amplitudes is difficult to see. In a recent paper \cite{Geyer:2015bja}, this difficulty was overcome for 10d model by transforming the 1-loop amplitude to a sphere, and it would be very interesting to implement a similar mapping of our 4d 1-loop amplitude. Although the loop integrands of the 4d ambitwistor model contain elliptic functions, they may still be useful for studying various properties of $\mathcal{N}=8$ supergravity. For example, since a genus $g$ amplitude
of the 4d ambitwistor string model should encode all $g$-loop Feynman diagrams of $\mathcal{N}=8$ supergravity with a given number of external legs, it should exhibit better power counting behavior than individual Feynman diagrams. Ultimately, we hope that the 4d ambitwistor model reveals new symmetries of $\mathcal{N}=8$ supergravity which provide insight into its possible finiteness.

{\it Acknowledgements:}
We thank Rutger Boels, Lionel Mason, and especially David Skinner for helpful discussions. This 
work was supported by the German Science Foundation (DFG) within the Collaborative Research 
Center 676 Particles, Strings and the Early Universe and by the People Programme (Marie Curie 
Actions) of the European Union's Seventh Framework Programme FP7/2007-2013/ under REA Grant
Agreement No 317089 (GATIS).

\end{document}